\documentclass[letterpaper]{article} 
\usepackage{aaai23}  
\usepackage{times}  
\usepackage{helvet}  
\usepackage{courier}  
\usepackage[hyphens]{url}  
\usepackage{graphicx} 
\usepackage{natbib}  
\usepackage{caption} 
\usepackage{algorithm}
\usepackage{algorithmic}
\usepackage{amsmath}

\usepackage{booktabs}
\usepackage{newfloat}
\usepackage{listings}
\DeclareCaptionStyle{ruled}{labelfont=normalfont,labelsep=colon,strut=off} 
\floatstyle{ruled}
\newfloat{listing}{tb}{lst}{}
\floatname{listing}{Listing}
\title{YouNICon: YouTube’s CommuNIty of Conspiracy Videos}
\author{
    Liaw Shao Yi$^1$, Fan Huang$^2$, Fabricio Benevenuto$^3$, Haewoon Kwak$^2$, Jisun An$^2$
}
\affiliations{
    $^1$School of Computing and Information Systems, Singapore Management University, Singapore\\
    $^2$Luddy School of
    Informatics, Computing, and Engineering, Indiana University Bloomington, USA\\
    $^3$Federal University of Minas Gerais Belo Horizonte, Brazil

    shaoyi.liaw.2022@phdcs.smu.edu.sg, huangfan@acm.org, fabricio@dcc.ufmg.br, 
    haewoon@acm.org, jisunan@acm.org
}

\begin{document}

\maketitle

\begin{abstract}
Conspiracy theories are widely propagated on social media. 
Among various social media services, YouTube is one of the most influential sources of news and entertainment. 
  This paper seeks to develop a dataset, \textsc{YouNICon}, to enable researchers to perform conspiracy theory detection as well as classification of videos with conspiracy theories into different topics. \textsc{YouNICon} is a dataset with a large collection of videos from suspicious channels that were identified to contain conspiracy theories in a previous study~\cite{Ledwich2020}. 
  Overall, \textsc{YouNICon} will enable researchers to study trends in conspiracy theories and understand how individuals can interact with the conspiracy theory producing community or channel. 
  Our data is available at: \url{https://doi.org/10.5281/zenodo.7466262}.
\end{abstract}

\section{Introduction}
Conspiracy theories are nothing new in human history. Scholarly research on conspiracy theories began in the 1930s \cite{Butter2018}, and it has been a field that is highly multidisciplinary and diverse~\cite{Mahl2022}. Various researchers have proposed definitions for conspiracy theories.  \citet{keeley1999conspiracy} defines conspiracy theory as ``a proposed explanation of some historical event (or events) in terms of the significant causal agency of a relatively small group of persons--the conspirators—acting in secret.'' 
A more general definition of conspiracy theory is provided by \citet{wood2012dead} as ``a proposed plot by powerful people or organizations working together in secret to accomplish some (usually sinister) goal.''

Conspiracy is sometimes considered a form of misinformation. Misinformation is commonly defined as ``false or inaccurate information that ... spread regardless of an intention to deceive.~\cite{10.1145/3460231.3474241}'' 
This suggests that any malicious intent of the content creator is not a necessary condition of misinformation, but incorrectness of information is. 
Thus, there is a stark difference between conspiracy and misinformation; the intent of powerful people (or organizations) is crucial for the definition of conspiracy. 
This difference proves the need for in-depth research on conspiracies that should be differentiated from those on misinformation. 

A belief in conspiracy often correlates with anomia, lack of interpersonal trust, and having political beliefs at extreme ends of the political spectrum (especially on the right-hand extreme)~\cite{10.2307/3791630, Sutton2020}. 
Conspiracy theories, in contrast to non-conspiracy views, tend to be more attractive as they satisfy one's epistemic (e.g., the desire for understanding, accuracy, and subjective certainty), existential (e.g., the desire for control and security), and social desires (e.g., the desire to maintain a positive image of the self or group)~\cite{douglas2017psychology}. 
This results in undesirable outcomes like decreased institutional trust and social engagement, political disengagement, prejudice, environmental inaction, and an increased tendency towards everyday crime~\cite{Pummerer2021,douglas2017psychology,Jolley2019}.
Additionally, conspiracy theories can form a worldview in which believers of a type of conspiracy tend to approve of other conspiracies as well \cite{ wood2012dead,dagnall2015conspiracy}. Polls have also shown that ``everyone believes in at least one or a few conspiracy theories~\cite{uscinski2020conspiracy}''. Hence, a holistic understanding of conspiracy theories cannot be achieved in isolation from a specific type of conspiracy.

In today's context, conspiracy theories are widely propagated on  social media~\cite{Enders2021,Mahl2021}. 
Conspiracy narratives are nourished by information cascades on social media and reach a larger audience \cite{Monaci2021}. 
Consequently, these false narratives tend to outperform real news in terms of popularity and audience engagement within online environments \cite{David2021,Vosoughi2018}. 
\citet{Enders2021} show that usage of 4chan/8kun has the highest correlation with the number of conspiracy beliefs, followed by Reddit, Twitter, and YouTube. 

Among the social media services, YouTube is one of the most influential sources of news and entertainment~\cite{pew}. 
It has 2,562 million monthly active users, and it is the second most popular social network worldwide as of January 2022~\cite{Statista1}, contributing to a billion hours of video viewed daily~\cite{Goodrow2017}.  
Audit studies show that video recommendations on YouTube can lead to the formation of filter bubbles on misinformation topics \cite{Hussein2020}. 
Similarly, exposure to conspiracy videos might result in undesirable outcomes. For example, the belief that the 5G cellular network caused COVID-19 has resulted in more than 200 reports of attacks against telecom workers in the United Kingdom \cite{vincent2020something}. The belief in white genocide conspiracies resulted in the death of 51 individuals in New Zealand \cite{shooting}. 
The belief in conspiracy theories is no doubt an issue of concern. 
However, most existing research focuses only on specific types of conspiracy theories, and not all datasets are available for research communities. 

In this work, we build \textsc{YouNICon}, a curated dataset of YouTube videos from channels identified as producing conspiracy content by Recfluence~\cite{Ledwich2020}. 
We aim to help researchers to study the patterns of production and consumption of conspiracy videos, such as how individuals interact with those videos from an aggregated (video) or individual level (comments)\footnote{\label{note1}The comment text and real author names are not shared to protect the identity of the commenter. The actual comment text can be rehydrated using YouTube Data API.}.

\textsc{YouNICon} comprises the following information: 
\begin{itemize}
\item Metadata of all 596,967 videos from 1,912 channels that produced conspiracy identified by Recfluence~\cite{Ledwich2020} 
\item A list of 3,161 videos manually labeled as being about conspiracy or not 
\item 37,199,252 comment IDs of comments in all videos with basic metadata and scores from the Perspective API\textsuperscript{\ref{note1}}
\item 100 videos manually labeled for the type of conspiracy. 
\end{itemize}

\textsc{YouNICon} will be a valuable resource for studying YouTube as a medium of conspiracy theory production and consumption. The contributions of this paper are as follows: 
\begin{itemize}
\item Curate a large-scale dataset of videos with conspiracy content (\url{https://doi.org/10.5281/zenodo.7466262})
\item Perform exploratory analyses on the dataset to understand its key properties
\item Discuss potential uses for the dataset
\end{itemize}

\section{Related Work and Datasets}
\subsection{Conspiracy Detection}
\begin{table*}[!ht]
    \centering
    \caption{Related Datasets for Conspiracy Detection}
    \label{tab:datasets}
    \begin{tabular}{p{3cm}p{6cm}p{4cm}p{3cm}}
    \toprule
        Dataset & Description & Labels  & Annotation Method  \\ 
    \midrule
        Recfluence \cite{Ledwich2020} & Political inclination of YouTube channels and tags to characterize channel. & 7,085 channels with 2,365 of the channels with the tag conspiracy & Agreement between 3 labelers  \\ 
    \midrule
         Conspiracy theory videos \cite{faddoul2020longitudinal} & YouTube Video Dataset of Conspiracy Theory Videos from YouTube's Watch Next engine & Conspiracy (542) and non-conspiracy (568) in training set & Conspiracy Videos from a Book \cite{jackson2017top} or Reddit and non-conspiracy videos from random scraping   \\ 
    \midrule
         Conspiracy theory Tweets and videos \cite{ginossar2022cross} & Cross-platform dataset which includes YouTube videos from Tweets related to COVID-19 and vaccines & 930,539 Tweets and 1,280 YouTube conspiracy theory videos with transcripts & Not applicable  \\ 
    \midrule
        Twitter conversations conspiracy \cite{marchal2020coronavirus}  & Twitter conversations dataset of conversations with more than 1000 Tweets that contain the word ``conspiracy". & More than 4,500 conversations & Semi-automatic   \\ 
    \midrule
        Hoaxes and Hidden agendas  \cite{Phillips2022} & Tweets from 4 Topics: climate change, COVID origins, COVID vaccine, and Epstein.  & 3,100 annotated instances. Conspiracy (2,336) and non-conspiracy (764)  & Agreement between 3 labelers.  \\    
    \midrule
        
        CMU-MisCOV19  \cite{memon2020characterizing}& Tweets related to COVID-19 misinformation with 17 different labels which include topics like Irrelevant, Conspiracy, True Treatment, Politics, Commercial Activity, etc.  & 4,573 annotated Tweets. Conspiracy (924)  & Annotation class determined by 1 single annotator.  \\ 
    \midrule
       \citet{Moffitt2021} & Extension of the dataset by  \citet{memon2020characterizing} using the same procedure, then collapsed labels into a binary conspiracy classification task.  & 8,781 labeled Tweets 4,573 Tweets from   \citet{memon2020characterizing} and 4,208 new labeled tweets. & 
       Manually labeled by research assistants. \\
    \bottomrule
    \end{tabular}
\end{table*}

Table~\ref{tab:datasets} highlights several existing datasets that have been used for conspiracy theory detection research. 
Existing literature often focuses on  misinformation~\cite{10.1007/978-3-030-32236-6_30,kumar2020fake} or 
specific conspiracy theories related to COVID-19, alien visitation, anti-vaccination, white genocide, climate change, or Jeffery Epstein~\cite{Moffitt2021, RR-A676-1, Phillips2022}. Most works focus mainly on Tweets as the unit of the study~\cite{Moffitt2021, marchal2020coronavirus, Phillips2022, Mahl2021}. 
For example, \citet{marchal2020coronavirus} study Tweets explicitly containing the word ``conspiracy.'' \citet{Phillips2022} have compiled a dataset consisting of four types of conspiracy, namely climate change, COVID-19 origins, COVID-19 vaccine, Epstein Maxwell.  
\citet{Moffitt2021} study COVID-19-related conspiracies by training a BERT-based classifier to distinguish conspiracy Tweets.

\subsection{Conspiracy Taxonomy}
\citet{Mahl2021} used network analysis of co-occurring hashtags in Tweets to assign hashtags into topic groups qualitatively based on their thematic relationship. This resulted in the 10 most visible conspiracies, which include Agenda 21, Anti-Vaccination, Chemtrails, Climate Change Denial, Directed Energy Weapons, Flat Earth, Illuminati, Pizzagate, Reptilians, and 9/11 Conspiracies. 
While co-occurring patterns of hashtags reveal a partial taxonomy of conspiracy, a more comprehensive one is found on Wikipedia.

On Wikipedia, a list of conspiracy theories is constantly being updated~\cite{enwiki:1095125436}. 
Upon closer inspection of the list of conspiracy topics from Wikipedia, we found that it covers well the conspiracies in \citet{Mahl2021} (see  Table \ref{tab:topics} for details). Hence, we will use the taxonomy of Wikipedia for \textsc{YouNICon}.

\section{Data Collection}

On YouTube, interactions between content creators and consumers occur as follows: a content creator posts a video with a title, description, and tags. A content consumer views, likes, or comments on a video. A ``view'' represents a  playback of a video, a ``like'' is positive feedback to the video by users, and a ``comment'' is the way in which online collective debates grow around the video~\cite{Bessi2016,YouTube2022}. 
A comment can be a reply to a video (a top-level comment) or a reply to other comments.

\subsection{YouTube Channels about Conspiracy}

\citet{Ledwich2020} curated a list of US-based political channels in the Recfluece project. They classify each channel based on its political leaning, channel type (e.g., mainstream news, AltRight, etc), and topical category (e.g., conspiracy, libertarian, organized religion, LGBT, etc). 

We downloaded an entire list of YouTube channels from Recfluence on 25 February 2022 and extracted only channels with the ``conspiracy'' label. 
We then used the YouTube Data API\footnote{\url{https://developers.google.com/youtube/v3}} to collect the basic information about these channels. Out of the 2365 channels with the ``conspiracy'' label, 1912 channels were accessible by the YouTube API. The rest of the channels were deleted from YouTube and thus excluded from the following analysis. 
While Recfluence provides a quite  extensive list of US-based political channels, the resulting list could be improved with more channels. However, all the pipelines used in this work will still be valid.

\subsection{Video Metadata}
In contrast to Recfluence~\cite{Ledwich2020} that provide channel-level conspiracy information, \textsc{YouNICon} focuses on video-level conspiracy. 
For the extracted conspiracy-related channels from Recfluence, we collect the metadata of every video published on those channels. 
The metadata includes the title,  description, tags, number of likes, number of views, duration, and  published date. 
We collect these metadata for 1,049,413 videos in total. 
To get a better sense of the content that was presented in the videos, we also collect transcripts or subtitles of the videos using a PyPI package, youtube-transcript-api\footnote{\url{https://github.com/jdepoix/youtube-transcript-api}}.  
We only consider those videos with English transcripts, which are 761,565 videos in total.

We further filter out non-English videos by detecting the language of the videos based on their titles, which often summarizes the gist of the video.  
In particular, we use the Fasttext language identification model~\cite{joulin2016bag}, which can recognize 176 languages, with a threshold of 0.5 to determine the language with the highest probability for a video. Between the two Fasttext language identification models, we used the larger and more accurate one (i.e., lid.176.bin). 
As a rule of thumb, channels with less than 80 percent of their videos that are in English are excluded from the rest of the analysis. 

For all textual metadata, we apply common preprocessing techniques (e.g., remove emojis, URLs, punctuation, and numbers, and convert them to lowercase).
Then, we filter out videos that do not have all the metadata. This results in a collection of 596,967 videos with all metadata, which are title,  description, tag, and transcript.

Additionally, we collect top-level comments as a part of the video's features. 
We filter out comments' authors if 1) their comments detected as English are less than 80\%, or 2) they leave only one comment.  
We also eliminate the top-level comments written by the same video creator to focus on the behavior of the viewers. 
As a result, we obtain 37,199,252 comments. 
For these comments, we use the Perspective API to perform scoring for toxicity, identity attack, and threat\footnote{\url{www.perspectiveapi.com}}.

\section{Dataset Construction}

\begin{figure}[th!] 
 \centering 
 \includegraphics[width=0.98\columnwidth]{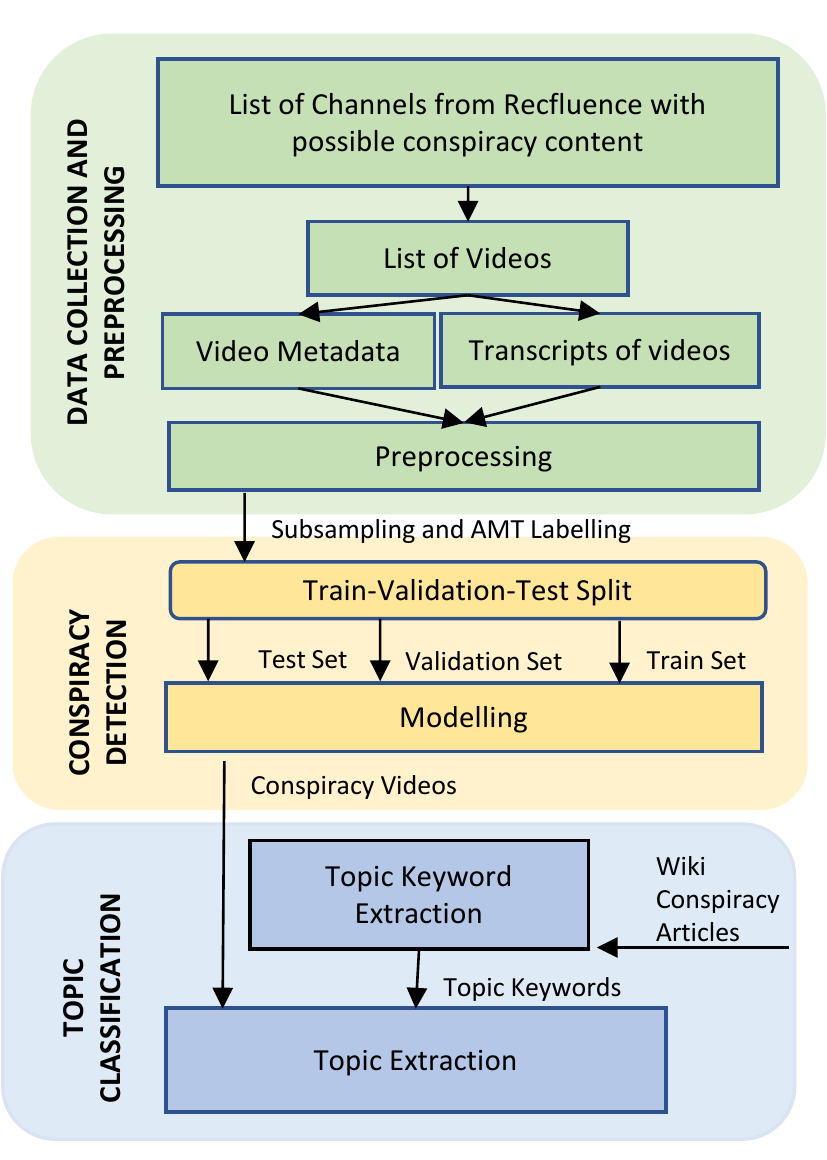} 
\caption{Overview of \textsc{YouNICon} Construction} 
\label{fig:pipeline}%
\vspace{-1em} 
\end{figure}

Figure~\ref{fig:pipeline} is a flowchart that summarises the dataset construction proposed in this paper. The following sections will explain the proposed method in detail. Table \ref{tab:variable_names} summarises the variables available in the \textsc{YouNICon}.

\begin{table}[!ht]
    \caption{Variables and descriptions in the \textsc{YouNICon} dataset}
    \label{tab:variable_names}
    \centering
    \begin{tabular}{p{2.5cm}p{5cm}}
        \toprule
        Variable & Description \\  \toprule
        \multicolumn{2}{c}{All Videos} \\
        \midrule
        Video\_id & YouTube Video ID \\ 
        Channel\_ID & ID of the channel that video is from \\ 
        Title & Title of the video \\ 
        Video\_Description & Description of Video Provided by content creators \\ 
        Tags & Tags provided by content creators \\ 
        Transcript & transcript of the video from the youtube-transcript-api \\ 
        Published\_date & date video is published \\ 
        Views & number of views \\ 
        Likes & number of likes \\ 
        Dislikes & number of dislikes \\ 
        NumComments & number of comments \\ 
        Duration & duration  \\ 
        Category & content category provided by YouTube \\ 
        DurationSec & duration of videos in seconds \\  
        \midrule
        \multicolumn{2}{c}{Comments} \\
        \midrule
        Comment\_Id & ID of comments \\ 
        VideoId & ID of the video comment is on \\ 
        Anon\_id &  anonymised author's ID \\ 
        Toxic & Toxic Score from Perspective API \\ 
        Identity & Identity Attack Score from Perspective API \\ 
        Threat & Threat Score from Perspective API \\ 
        LikeCount & number of likes on comment \\ 
        PublishedAt & publish time of comment \\ 
        TotalReplyCount & number of replies to comment \\  
        \midrule
        \multicolumn{2}{c}{Conspiracy Label} \\
        \midrule
        Video\_id & YouTube Video ID \\ 
        Majority\_label & 1 if the video contains conspiracies 0 otherwise \\  
        \bottomrule
    \end{tabular}
\end{table}

\subsection{Video Labeling}
The procedure of manual annotation is done in accordance with the Institutional Review Board (IRB) guidelines under the approval number IRB-22-129-A071(922) of Singapore Management University. 

We use Amazon Mechanical Turk (AMT) for data labeling. Our labeling task, known as Human Intelligence Task (HIT) in AMT, asks an AMT worker whether a given video contains conspiracy or not. The title, description, tags, and the first 1,000 characters of the transcript of each video are given to AMT workers. 
We select workers located in the US, with a past HIT approval rate of greater than 98\% and 5,000 HITs approved, and compensated them at a rate of 0.05 USD per HIT.

For each video, we recruit three workers and determined a label based on the majority vote. 
In contrast to misinformation where there is a clear-cut answer, determining whether a video contains conspiracy can be more challenging as an individual's political or religious belief might affect their decision about conspiracy videos. Thus, we follow a majority voting scheme for each video's label.

Labeling is conducted in two stages. In the first stage, we sample 2,200 videos. After labeling, we find that this dataset is somewhat imbalanced; only around 20\% (436 of the videos out of 2,184) contain conspiracy theories. Although 20\% may seem like a relatively large proportion of the videos with conspiracy theories, we note that all these videos are from the channels that are categorized as `conspiracy' in Recfluence~\cite{Ledwich2020}. 
To make \textsc{YouNICon} a better-balanced dataset of conspiracy and non-conspiracy videos, we use Machine Learning models to get pseudo-labels first. We finetune the RoBERTa-large model by using the sampled videos. 
We split the data into train, validation, and test sets and use the concatenated texts as features for the model. 
This model attained an accuracy of 0.74, with a positive F1 of 0.5273 and a negative F1 of 0.8207. 
We used this trained model to assign `conspiracy' or `non-conspiracy'  pseudo-labels to all the videos in  the full dataset. 
Then, we sample 1,000 videos with `conspiracy' pseudo-labels and manually labeled them in the same manner. 
After these 2 rounds of labeling, we obtain a dataset of 3,161 videos (1,144 conspiracy videos (36.2\%)). 
Fleiss' Kappa, an extension of Cohen's kappa, is used to measure inter-rater reliability~\cite{fleiss1971measuring}. A score of 0.4111 is calculated, implying that there is a moderate agreement between raters in the dataset~\cite{landis1977measurement}.

\begin{table*}[!ht]
    \centering
    \caption{Topics of Conspiracy Theories and the corresponding keywords}
    \label{tab:topics}
    \begin{tabular}{p{3.9cm}p{5.5cm}p{7cm}}
    \toprule
        Topic & Description & Representative Keywords  \\ 
        \midrule
        Aviation & Chemtrails, Air travel and aircraft & chemtrail, black helicopter, airline, aircraft, \newline airlines, remotely, flight, crash  \\ 
        Business and Industry & Deep Water Horizon and New Cocacola formula &  cocacola, deepwater, coke, formula  \\ 
        Deaths and Disappearances & Deaths of prominent leaders and public \newline figures & jfk, dnc, flee, lookalike, assassination  \\ 
        Economics and Society & New World Order, George Soros, \newline Freemasonry & masonic, new world order, george soros, turkey, freemasonry, üst akıl, mastermind, denver airport, economy, freemason, erdoğan, rip  \\ 
        Espionage & Spying with animals or individuals &  animal, taliban, spy, wilson, malala, harold, \newline golitsyn  \\ 
        Ethnicity, Race, and Religion & Related to Anti-Religion, Racism, \newline Genicides, and Religious Beliefs & antisemitism, jesus, antichrist, paul, rastafari, catholicism, islamic, bah\'a'\'i, bible, apostle, racism, christ, islam, catholic, jihad, bah\'a'\'ism, armenianism  \\ 
        Extraterrestrials and UFOs & Alien visitation & ufo, anunnaki, extraterrestrial, alien  \\ 
        Government, Politics, and Conflict & 9/11. False Flag operations, Political \newline Figures & election, congress, trump, marxism,barack, \newline clinton, obama, epstein, ukraine, sandy hook, biden, clintons, national, crisis actor, fema  \\ 
        Medicine & COVID-19 related, Claims that diseases like HIV and Ebola is invented, \newline anti-vaccination, Water fluoridation &  vaccination, fluoridation, vaccine, disease, \newline therapy, suppression, virus, pandemic, \newline pharmaceutical, virology  \\ 
        Science and Technology & Global Warming Denial, Flat Earth, Weather Control, Technology Surpression & rfid, weather, earthquake, weaponry, weather \newline control, mkultra, tsunami, haarp, mind control, warming, earth, warm, technology, flat  \\ 
        Outer Space & Staged moon landings by NASA, Nibiru (doomday belief of large planet almost crashing on Earth) & nibiru, outer space, planet, nasa, solar, space \\ 
        \bottomrule
    \end{tabular}
\end{table*}

\subsection{Topic Classification} \label{method taxonomy}

Going beyond whether a video is about conspiracy or not, we also assign a conspiracy topic to a video based on conspiracy taxonomy compiled on Wikipedia~\cite{enwiki:1095125436}.
In doing so, we first parse the text of the ``List of conspiracy theories'' page on Wikipedia~\cite{enwiki:1095125436}. 
This Wikipedia page contains summaries of the popular conspiracy theories, which include Aviation, Business and Industry, Deaths and Disappearances, Economics and Society, Espionage, Ethnicity, Race and Religion, Extraterrestrials and UFOs, Government, Politics and Conflict, Medicine, Science and Technology, Outer Space and Sports (Table \ref{tab:topics}). We exclude the category of ``Sports'' because our dataset, based on Recfluence~\cite{Ledwich2020}, is unlikely to contain Sports-related conspiracies. 
The topic ``Fandom, celebrity relationships, and shipping'', a new topic added on 18 May 2022, which is after our Wikipedia data collection, is also not included in this analysis.

The topic classification consists of two stages: 1) keyword extraction and 2) topic inference. 
For keyword extraction, we identify  representative words of each topic using log-odds ratios with informative Dirichlet priors \cite{monroe2008fightin}, which is a widely used technique for a large-scale comparative text analysis~\cite{an2021predicting,kwak2020systematic}. It estimates the log-odds ratio of each word between two corpora $i$ and $j$ given the prior frequencies obtained from a background corpus. 
We rank the words based on their log-odds scores and obtain a list of representative words for each of the conspiracy theories. The background corpus used in this analysis is the ``google 1-gram'' \cite{michel2011quantitative}, extended with the counts of the vocabulary used in the ``list of conspiracy theories'' Wikipedia page. 
For each conspiracy topic, we compare the corpus of one topic against the concatenated corpus of all other topics. 
For each topic, we use the top ten keywords as the preliminary keywords for topic inference (see Table \ref{tab:topics} for the list of words extracted). 
We also add the subtopic names listed in Wikipedia to the keywords of each topic. 
We then convert 21 keywords to bigrams or trigrams to be more distinguishable (e.g., the keywords new, world, and order should be considered as a trigram, not three unigrams) and remove 62 keywords related to countries or locations (Malaysia, Wuhan) or those that are generic (January, human).

Having the representative keywords for topics at hand, we infer the topic of the video by simply using a keyword-matching method. We match the keywords in each topic to the video's features by using spaCy's PhraseMatcher. We assign a topic by choosing one with the highest frequency of keywords in a video's features.

\begin{table}
    \centering
    \caption{Feature comparisons between conspiracy (C) and non-conspiracy (NC) videos. Avg. and Med. are average and mean values,  respectively.}
    \begin{tabular}{p{2cm}p{1cm}p{1.2cm}p{1cm}p{1.2cm}}
    \toprule 
        \textbf{Feature} & Avg.(C) & Avg.(NC) & Med.(C) & Med.(NC)  \\ 
    \midrule
        \textbf{Duration (s)} & 1,398 & 1,298 & 688 & 638  \\ 
        \textbf{Likes}  & 736 & 645 & 136 & 72  \\ 
        \textbf{Comments} & 171 & 113 & 32 & 14  \\ 
        \textbf{Views} & 27,245 & 20,021 & 4,317 & 2,018 \\ 
    \bottomrule
    \end{tabular}
    \label{mann}
\end{table}

\subsection{Exploratory Data Analysis}

To provide a brief overview of the dataset, we conduct an exploratory analysis.  We first compare the difference in engagement of videos with conspiracy theories and those without conspiracy theories. 
Table~\ref{mann} shows the average and median values of various features of conspiracy and non-conspiracy videos. 
Conspiracy videos have longer lengths and get more likes, comments, and views than non-conspiracy videos. Conspiracy videos have 736 likes, 171 comments, and 27,245 views on average, while non-conspiracy videos have only 645 likes, 113 comments, and 20,021 views. 
All differences are statistically significant based on the Mann Whitney U test \cite{mann1947test} for unpaired samples.

\begin{table*}[!ht]
    \centering
    \caption{Conspiracy detection result of baselines and RoBERTa-based model (R) with different features}
    \begin{tabular}{l|cccccc}
    \toprule
        Model & accuracy & recall & precision & F1 weighted & F1 negative & F1 positive \\ \midrule
        Dummy Classifier & 0.8000 & 0.0000 & 0.0000 & 0.7111 & 0.8889 & 0.0000 \\ 
        Naïve Bayes & 0.6825 & 0.8125 & 0.3672 & 0.7141 & 0.7661 & 0.5058 \\ 
        Logistics & 0.7750 & 0.7750 & 0.4627 & 0.7930 & 0.8464 & 0.5794 \\ 
        SVM Linear & 0.7675 & 0.7625 & 0.4519 & 0.7863 & 0.8410 & 0.5674 \\ 
        \midrule
        combined (R) & \textbf{0.8575} & 0.7500 & \textbf{0.6186} & \textbf{0.8624} & \textbf{0.9085} & \textbf{0.6780} \\ 
        transcript (R) & 0.7875 & \textbf{0.8250} & 0.4818 & 0.8050 & 0.8542 & 0.6083 \\
        video description (R) & 0.8075 & 0.7000 & 0.5138 & 0.8177 & 0.8740 & 0.5926 \\ 
        Title (R) & 0.8025 & 0.6875 & 0.5046 & 0.8130 & 0.8707 & 0.5820 \\ 
        Tags (R) & 0.8100 & 0.6875 & 0.5189 & 0.8193 & 0.8762 & 0.5914 \\ 
        \bottomrule
    \end{tabular}
    \label{result_test}
\end{table*}

\section{Results}
\subsection{Conspiracy Detection}

We use the annotated data of 3,161 videos to build a classifier that detects whether a video is about conspiracy or not. Since our data is slightly unbalanced (1,144 videos are conspiracy), we perform under-sampling to balance the classes for the training. For testing, we use the holdout test set sampled from the initial (first-round) 2,200 annotated videos. 

As a feature, we use all video's textual meta information, including title, tags, description, and transcript. 
Since the deep learning models can take 512 tokens at maximum~\cite{Liu2019}, we truncate the video description and transcript, using the first 200 tokens. The feature input, called combined, is created by concatenating the first 200 tokens or words for both the video description and transcript, followed by the title and tags.

In order to compare the performance of the models, other than simply accuracy, recall, or precision, we use the F1-score: 
\[
F_{1}=\frac{2 \times \text{Precision} \times \text{Recall}}{\text{Precision} + \text{Recall}}
\], which is calculated for both the positive and negative classes. To account for class imbalance, F1 weighted, which is the F1 score weighted by the support, is also used. 

Table~\ref{result_test} summarizes the prediction results of various models. 
Dummy Classifier predicts all videos as negative (or non-conspiracy), yielding an accuracy of 0.8, which is the same as the proportion of non-conspiracy videos in the test set. 
Traditional machine learning models, including Naive Bayes, Logistics Regression, and Support Vector Machine with Linear Kernel (SVM), are also tested.
All three models, Naive Bayes, Logistics, and SVM slightly outperform the Dummy classifier, obtaining a weighted F1 of 0.7141, 0.7930, and 0.7863, respectively. 

We also explore pre-trained language models, such as RoBERTa-large. 
The training set is further split into 80-20 train-validation split for finetuning of the pre-trained model. The learning rate of 1e-5 is used with a batch size of 4 with random seed 13 for finetuning. 
Our results in Table~\ref{result_test} show that the pre-trained models result in better performance in all metrics but recall, obtaining an accuracy of 0.8575 and weighted F1 of 0.8624. 

In Table~\ref{result_test}, we also show the prediction results based on individual features.  
By comparing the weighted F1 of models built based on each feature, we observe that the tags are best among the individual features, followed by video description, titles, and transcript.

\subsection{Zero Shot and Few Shot Classification}

We further conduct experiments to examine if it is possible to detect conspiracy theories via zero and few-shot learning. Zero and few-shot learning are techniques that aim to make predictions for new classes with limited labeled data. 
We test pre-trained Natural Language Inference (NLI)~\cite{bowman-etal-2015-large, williams-etal-2018-broad} and Natural Language Generation (NLG)~\cite{lewis-etal-2020-bart, zhang2022opt} models on zero and few-shot settings. We use all the features of videos as the input for those models.

For the NLG models, we test on both auto-regressive generation and sequence to sequence models\footnote{opt-125m for auto-regressive model and bart-base for the sequence to sequence model}. 
However, we find that the generated results of zero-shot and most few-shot models are simply a repeat of the given text, from which we cannot infer the classification labels. 
The model could generate clear classification indicators (i.e., \textit{yes} or \textit{no} in our setting) only for the few-shot settings with 128 fine-tuning data instances. However, it predicts all inputs as non-conspiracy.

As for the NLI models, we apply the top three most popular fine-tuned zero-shot inference models from the Huggingface website\footnote{bart-large-mnil, distilbart-mnli-12-1, xlm-roberta-large-xmli}. 
Considering the NLI is not a binary classification task, we neglect the score of \textit{neutral} prediction and activate the scores of \textit{entailment} and \textit{contradiction} predictions as the final binary output.
To help the NLI models better understand the objective of detecting the conspiracy from short texts, we concatenate the input text with the assumption statement (i.e., \textit{That is a conspiracy.}). The model would then give out the answer about whether the assumption statement entails or contradicts the given text. The contradicting answer means that the model predicts the given text as non-conspiracy.
The zero-shot test results are in Table~\ref{zero_shot_NLI}. 
The best positive f1-score of 0.57 still does not outperform our proposed conspiracy detection method. Yet, we demonstrate the possibility of those NLI models for the conspiracy detection task.

\subsection{Topic Classification}

\begin{table}
    \centering
    \caption{Zero-shot and Few-shot F1-scores for NLI Models. The W stands for weighted f1-score; the N stands for Negative score, and the P stands for Positive score. For the settings column, the ZS stands for zero-shot, while FS-16 stands for Few-shot fine-tuned by 16 data instances.}
    \begin{tabular}{lcccc}
    \toprule
        Model & F1-W & F1-N & F1-P & Setting\\ \midrule
         & 0.83 & 0.91 & 0.53 & ZS\\ 
         & 0.83 & 0.91 & 0.54 & FS-16\\ 
        bart-large-mnli & 0.72 & 0.77 & 0.53 & FS-32\\ 
         & 0.82 & 0.88 & \textbf{0.57} & FS-64\\ 
         & 0.79 & 0.85 & 0.54 & FS-128\\ \midrule
         
         & 0.83 & 0.91 & 0.52 & ZS\\ 
         & 0.83 & 0.91 & 0.52 & FS-16\\ 
        distilbart-mnli-12-1 & 0.65 & 0.69 & 0.48 & FS-32\\ 
         & 0.77 & 0.83 & 0.52 & FS-64\\ 
         & 0.83 & 0.90 & 0.54 & FS-128\\ \midrule
         
         & 0.34 & 0.34 & 0.34 & ZS\\ 
         & 0.32 & 0.31 & 0.33 & FS-16\\ 
        xlm-roberta-large-xnli  & 0.40 & 0.42 & 0.32 & FS-32\\ 
         & 0.65 & 0.72 & 0.36 & FS-64\\ 
         & 0.62 & 0.68 & 0.37 & FS-128\\
    \bottomrule
    \end{tabular}
    \label{zero_shot_NLI}
\end{table}

We perform topic inference to understand the type of conspiracy theory of a video published on YouTube. 
For the ground-truth dataset, we randomly sample 100 videos and label them by the first author based on Table~\ref{tab:topics}. 

\begin{figure*}[th!] 
 \centering 
 \includegraphics[width=0.98\textwidth]{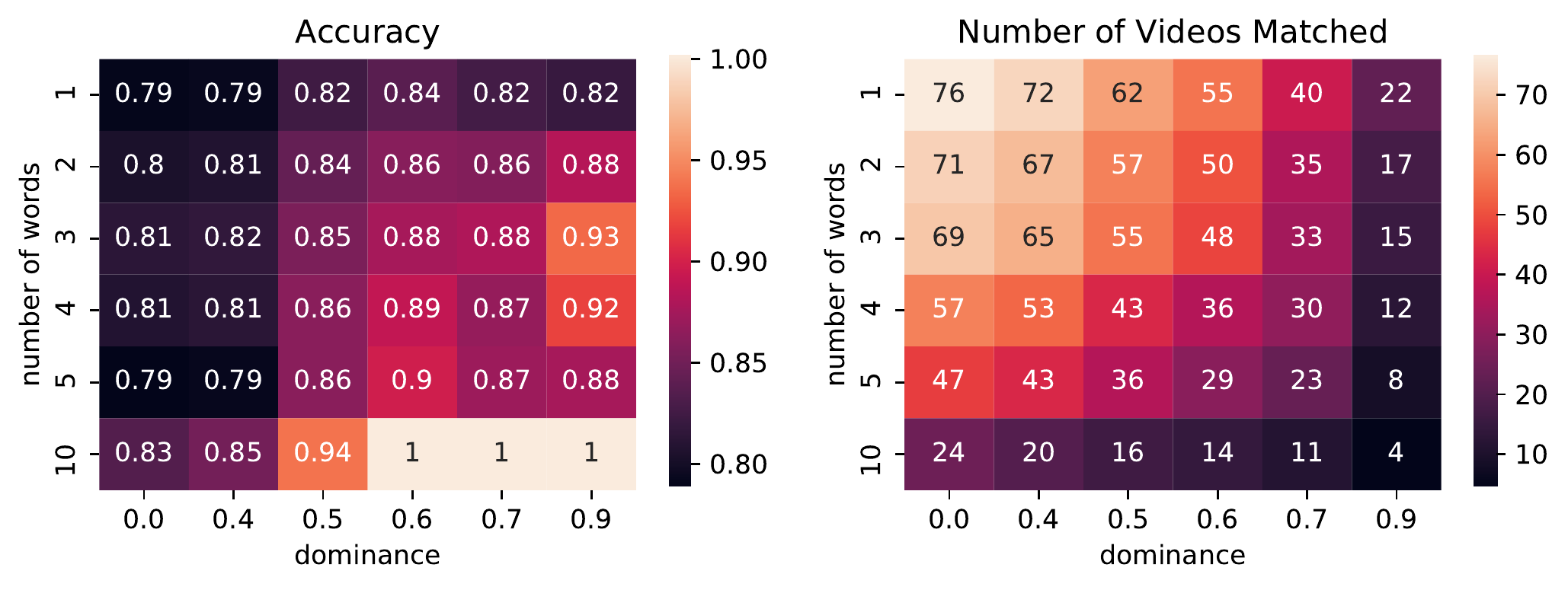} 
\caption{Sensitivity of Number of words and Dominance on Accuracy and Videos Matched} 
\label{fig:sensitivity}%
\vspace{-1em} 
\end{figure*}

In Figure~\ref{fig:sensitivity}, we investigate how sensitive our topic inference method is. 
The method has two parameters: dominance and the minimum number of words matched.
Dominance~\cite{zumpe1986dominance} is a metric that is commonly used to study the diversity of a community. 
A higher dominance score suggests a higher percentage of the words matched with one topic (i.e., if dominance is 1, all words matched are in one topic). 
Hence, having a threshold for dominance to be higher will ensure that the matched topic will have higher accuracy, but lesser videos are likely to be matched. 
The number of words matched also interplays with the retrieval and accuracy. 
Figure~\ref{fig:sensitivity} shows this relationship. 
For example, when we consider a match of the topic to be that it requires at least one word matched, 76 videos are matched with a topic, but the accuracy of matching is 0.789. 
If we increase the threshold to at least 10 words and the dominance score to be greater than 0.6, only 14 videos are matched with a topic but all matching is correct. 
When there are at least 2 words matched and a dominance threshold of greater than 0.5, the accuracy is 0.842, and 57 videos are matched with a topic.

\begin{figure}[h]
  \includegraphics[width=0.98\columnwidth]{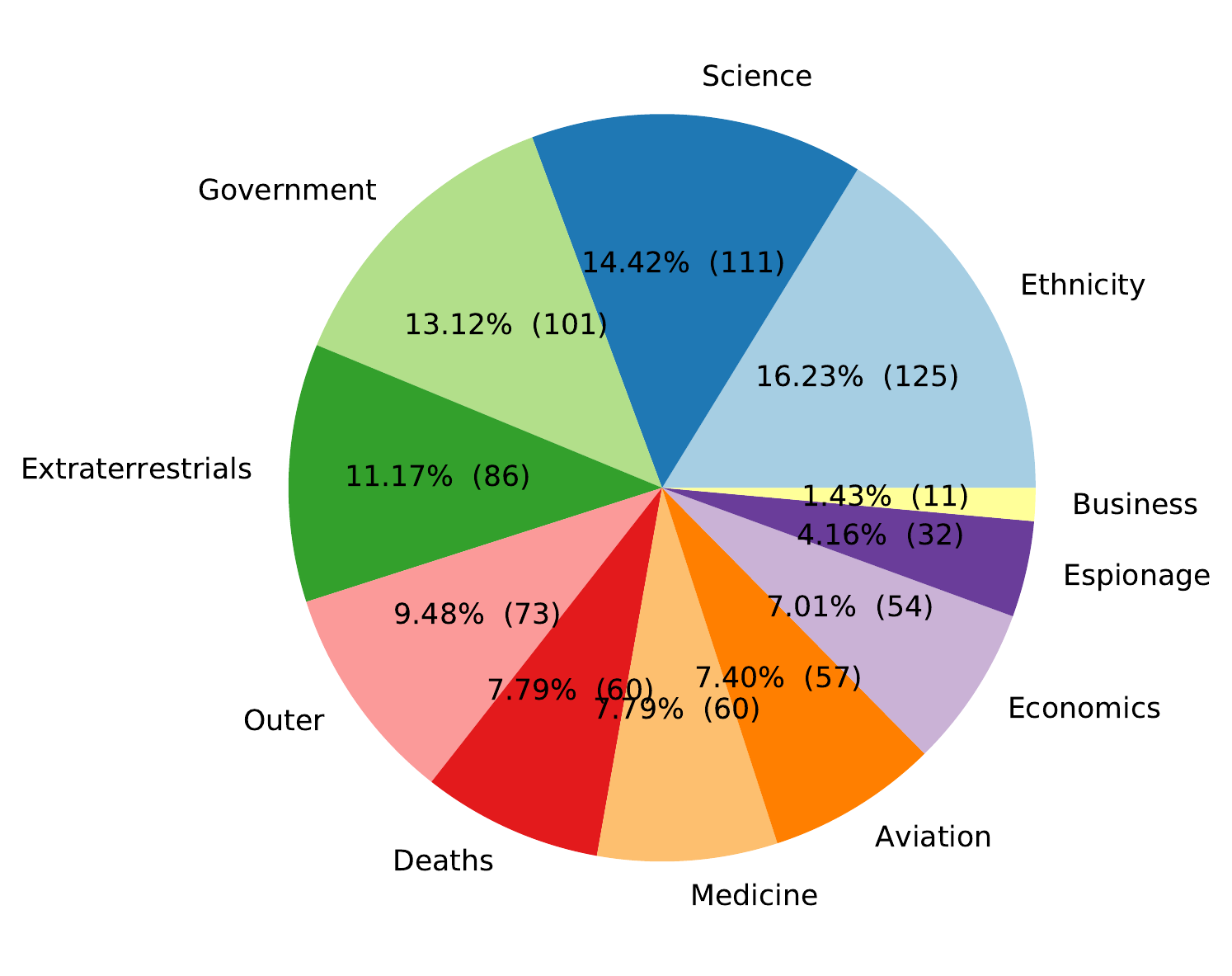} 
\caption{Topic distribution} 
\label{fig:TC}%
\end{figure} 

We explore the topics covered by conspiracy videos using the method outlined above. By using the parameters of at least 2 words matched and a dominance threshold of greater than 0.5, we apply the matching to all the videos with conspiracy theories in our dataset to understand the distribution of the conspiracy topics.
Out of the 1,144 conspiracy videos in the dataset, 770 videos have been matched with a topic. 
Figure~\ref{fig:TC} shows the distribution of the detected topics. 
Topics are relatively well distributed, and the top four topics are ``Ethnicity, Race, and Religion,'' ``Government, Politics, and Conflict,'' ``Science and technology,'' and ``Extraterrestrials and UFOs.''

\section{Discussion}
In this paper, we propose a new  dataset, \textsc{YouNICon}, for the detection of conspiracy theories on YouTube over various topics. While conspiracy theories have been studied for decades across different disciplines, a large-scale dataset of videos on popular social media services will accelerate research on the production and consumption of conspiracy theories on online platforms.

\textsc{YouNICon} offers a plethora of opportunities to study the subject of conspiracy theories from the text data. First, we hope that the automatic detection of conspiracy theories can be deeply explored by the machine learning community and potentially result in real-world tools to assist and facilitate the work of fact-checkers (e.g., pointing out not only conspiracy theories videos but the exact time the conspiracy theory appears within the video). Our study gives a first step in this direction by exploring standard classification techniques, providing the first assessment of the potential of automated detection of conspiracy theories, and also a baseline for future comparisons. Second,  
we hope researchers can use the dataset to study the dynamics of conspiracy theories on systems like YouTube. 
As this dataset contains all videos that are available in the channel's lifetime (as long as it is not removed from the platform), 
we are able to study how these content creators have evolved their production strategies over time. 
For example, do channels focus on a particular type of conspiracy over time or do they adopt a more generalist approach and produce a variety of content? Are there relationships between engagement and topics of conspiracy? The dataset has the potential to answer such questions. 

Similarly, for the content consumers (or the video audience), the comments included in the dataset can act as a peephole for us to analyze the behaviour of their consumption of conspiracy theories. In other words, researchers can potentially trace a conspiracy pathway, and look at how people can get involved in the echo chambers of conspiracy theories. 

Future works can include looking for better ways to perform topic classification. While Wikipedia’s list of conspiracy theories is used here, this classification can serve as a starting point for a better taxonomy to be developed. 
Given the advances of large language models (LLMs), it would be worth exploring the prompting approach with recent LLMs or the in-context learning approach with prompt tuning for the conspiracy detection task. 

\subsection{FAIR Consideration} 
The proposed dataset follows the FAIR principles of Findability, Accessibility, Interoperability, and Reuse-ability. The dataset can be found and accessed through Zenodo at the DOI: \url{https://doi.org/10.5281/zenodo.7466262}. 
Keywords for the topics of conspiracy theories are also shared as a CSV file for the use of other researchers for works related to conspiracy theories. Hence, the data satisfies reusability and interoperability.

\subsection{Ethical Consideration}
We carefully designed our dataset from the data collection period. We collect only publicly available data on YouTube with the use of YouTube's Data API. Also, our approach is approved by the Institutional Review Board of Singapore Management University (IRB-22-129-A071(922)). To safeguard the interests of our labelers on Amazon Mechanical Turks, they are informed that the content that the conspiracy theories are not true and that withdrawal from the study is without penalty. Helplines are also provided to the participants in the event of any negative emotions.  

\bibliography{aaai23}

\end{document}